\documentclass[12pt]{article}
\renewcommand{\baselinestretch}{2}
\usepackage{graphicx,amsmath,amssymb}
\begin{document}

\title{Magnetoelectronic states of a monolayer graphite\\}
\author{J H Ho, Y H Lai, Y H Chiu and M F Lin\\
\small  \it Department of Physics, National Cheng Kung University,
Tainan, Taiwan\\
\\
\small  \it E-mail address: mflin@mail.ncku.edu.tw}
\renewcommand{\baselinestretch}{1}
\date {}
\maketitle

\renewcommand{\baselinestretch}{1.5}
\begin{abstract}

The Peierl's tight-binding model, with the band Hamiltonian
matrix, is used to calculate the magnetoelectronic structure of a
monolayergraphite. There are many flat Landau levels and some
oscillatory Landau levels. The low Landau-level energies are
characterized by a simple relation, not for others. State
degeneracy is, respectively, fourfold degenerate and doubly
degenerate at low and high energies. The level spacing declines
quickly and then grows gradually in the increase of state energy.
The main features of electronic properties are directly reflected
in density of states. The predicted results could be verified by
the optical spectroscopy.

\vskip 0.6 truecm

\end{abstract}
\renewcommand{\baselinestretch}{2}

\newpage

Graphite, a layered system made up of carbon hexagons, is one of
the most extensively studied materials [1,2]. Recently, the
ability of controlling the film thickness in single-atom accuracy
has opened the new interest on the few-layer graphites [3]. Owing
to extraordinary electronic properties, the few-layer graphites
have attracted many studies, e.g., growth [4,5], band structure
[6,7], Coulomb excitations [8,9], optical spectra [10,11], and
transport properties [12,13]. A monolayer graphite is suitable in
studying the essential 2D quantum phenomena; furthermore, it has
the great potential in the nanoscaled electronic devices. This
system could exhibit rich physical properties, such as the
temperature-induced plasmon [8], the quantized absorption spectrum
[10], and the half-integer quantum Hall effect [13]. Such
properties are attributed to the peculiar band structure
associated with the underlying hexagon symmetry.

A single-layer graphite was a zero-gap semiconductor, mainly owing
to two linear bands just intersecting at the Fermi level
$E_F=0$[1]. Electronic properties were strongly affected by the
external fields. The spatially modulated electric and magnetic
fields were predicted to cause drastic changes in state
degeneracy, energy dispersion, band-edge state, and band width
[14,15]. The Landau levels due to a perpendicular uniform magnetic
field had been studied through the tight-binding model [6] and the
effective-mass approximation [2]. The former analyzed the field
effects on energy dispersion, energy spacing, band width, and
period of oscillatory Landau levels. However, those results were
obtained within a very large field $B>10^3$ T. The latter
predicted the formation of Landau levels only at low energies
($<1$ eV), so it was deficient in understanding energy dispersion,
state degeneracy, level spacing between two neighboring Landau
levels, and dependence of state energy on quantum number and
magnetic field. In this work, with the improvement on numerical
techniques, we could solve these problems and obtain
magnetoelectronic properties in a realistic field $B<50$ T. The
magneto-optical absorption spectra could be utilized to examine
the predicted results.

A monolayer graphite exists in a uniform perpendicular magnetic
field $\mathbf{B}=B\hat{z}$. The magnetic flux through a hexagon
is $\Phi=3\sqrt{3}Bb^{\prime 2}/2$ ($b^\prime=1.42{\AA}$ bond
length); it would be in unit of a flux quantum $\Phi_0=hc/e$. The
magnetic field induces the Peierl's phase characterized by the
vector potential $\mathbf{A}=Bx\hat{y}$. The nearest-neighbor
transfer integral related to the extra position-dependent phase is
\begin{align}
\langle
b_{j\mathbf{k}}|H_{\mathbf{B}}|a_{i\mathbf{k}}\rangle=\gamma_0\exp\{i[\,\mathbf{k}\cdot
(\mathbf{R}_i-\mathbf{R}_j)+\frac{2\pi}{\Phi_0}\int_{\mathbf{R}_i}^{\mathbf{R}_j}\mathbf{A}\cdot
d\mathbf{r}\,]\}.
\end{align}
$\gamma_{0}$(=2.56 eV[2]) is the atom-atom interaction between two
neighboring atoms at $\mathbf{R}_i$ and $\mathbf{R}_j$, and
${|a_{i\mathbf{k}}\rangle}$ ($|b_{j\mathbf{k}}\rangle$) is the
linear superposition of the $2p_z$ orbitals of the periodic $a_i$
($b_j$) atoms [Fig. 1]. The transfer integrals due to three
nearest-neighbor atom-atom interactions are given by
$t_{1\mathbf{k}}(n)=\gamma_{0}\exp[\,(ik_{x}b^\prime/2+ik_{y}\sqrt{3}b^\prime/2)+G_n\,]$,
$t_{2\mathbf{k}}(n)=\gamma_{0}\exp[\,(ik_{x}b^\prime/2-ik_{y}\sqrt{3}b^\prime/2)-G_n\,]$,
and $t_{3\mathbf{k}}=\gamma_{0}\exp(\,-ik_{x}b^\prime\,)$.
$G_n=i\pi\Phi[\,(n-1)+1/6\,]$. The magnetic flux would result in
the periodical boundary condition along $\hat{x}$ so that the
Peierl's phases are periodic in a period $2/\Phi$(=$2R_B$). The
rectangular primitive unit cell includes $4R_B$ atoms. These atoms
define the basis of the Hilbert space and can be used to represent
the Hamiltonian matrix. The order of the base functions is very
important here. If we naively arrange the base functions in an
order as a atomic label [Fig. 1], it would be very difficult in
diagonalizing a huge matrix[6]. The dimensionality is $4R_B>6400$
for $B<50$ T. To simplify the calculation problem, it is
convenient to choose the base functions in the following sequence,
$\{|a_{1\mathbf{k}}\rangle,|b_{2R_B\mathbf{k}}\rangle,
|b_{1\mathbf{k}}\rangle,|a_{2R_B\mathbf{k}}\rangle,\ldots
|a_{R_B\mathbf{k}}\rangle,|b_{R_B+1\mathbf{k}}\rangle,
|b_{R_B\mathbf{k}}\rangle,|a_{R_B+1\mathbf{k}}\rangle\}$. The
Hamiltonian is then represented as a $4R_B\times4R_B$ Hermitian
matrix,
\begin{align}
\left(
  \begin{array}{cccccccc}
    0 & q^{\ast} & p^{\ast}_1 & 0 & \ldots & \ldots & 0 & 0 \\
    q & 0 & 0 & p_{2R_B} & 0 & \ldots & \ldots & 0 \\
    p_1 & 0 & 0 & 0 & q^{\ast} & 0 & \ldots & 0 \\
    0 & p^{\ast}_{2R_B} & 0 & 0 & 0 & q & 0 & 0 \\
    \vdots & \ddots & q & 0 & 0 & \ddots & \ddots & 0 \\
    \vdots & \ldots & \ddots & q^{\ast} & \ddots & \ddots & 0 & p_{R_B+1} \\
    0 & \vdots& \vdots & \ddots & \ddots & 0 & \ddots & q \\
    0 & 0 & 0 & 0 & 0 & p^{\ast}_{R_B+1} & q^{\ast} & 0 \\
  \end{array}
\right),
\end{align}
where $p_n\equiv t_{1\mathbf{k}}(n)+t_{2\mathbf{k}}(n)$ and
$q\equiv t_{3\mathbf{k}}$. The range of $k_x$ is much smaller than
that of $k_y$ for $B<50$ T; it is sufficient to focus on energy
dispersions along $\hat{k_y}$. The unoccupied conduction band
($E^c$) is symmetric to the occupied valence band ($E^\nu$) at
$E_F=0$. Only the former is discussed in this work.

The $\pi$-electronic structure at $B=0$ exhibits two kinds of
energy dispersions and state degeneracies, as shown in Fig. 2(a).
Two nondegenerate linear bands just interact at $E_F=0$ or
$k_y=2\pi/3b^{\prime}$. Other bands are parabolic and doubly
degenerate. Figures 2(b)-2(d) at $B=40$ T show that the magnetic
field could make electronic states condense and produce Landau
levels. The number of magnetic subbands is inversely proportional
to $B$. Most of magnetic subbands are flat Landau levels except
some oscillatory Landau levels with very weak energy dispersions
at $E^{c}\sim\gamma_0$ [Fig. 2(c)]. Roughly speaking, a monolayer
graphite in the magnetic field could be regarded as a
zero-dimensional system. Each discrete Landau level could be
expressed as $E^c(n)$, where $n=0,1,2,\cdots$ is quantum number.
In addition to energy dispersion, the magnetic field also leads to
drastic changes in state degeneracy and level spacing. The Landau
levels with $E^c(n)<\gamma_0$ are fourfold degenerate , while
others are doubly degenerate. The level spacing ($E_s$) between
adjacent Landau levels is not equally spaced. $E_s$ is maximal
between the first two neighboring Landau levels, $E^c(0)$ and
$E^c(1)$ [Fig. 2(b)]. It decreases sharply with n increasing and
approaches to zero at $E^c\sim\gamma_0$ [Fig. 2(c)]. Then it grows
gradually in the further increase of n [Fig. 2(d)]. Both state
degeneracy and level spacing strongly depend on state energy;
therefore, a monolayer graphite is in sharp contrast to a 2D
electron gas [16].

The dependence of the Landau-level energies on n and B is quite
different from each other at low and high energies. The low Landau
levels with $E^c(n)\lesssim{}0.4\gamma_0$, as shown in Figs. 3(a)
and 3(b), are proportional to $\sqrt{n}$ and $\sqrt{B}$
simultaneously. However, there is no simple relation for the high
Landau levels with $E^c(n)\gtrsim{}0.4\gamma_0$. The previous
study by the effective-mass approximation predicts the simple
relation $E^c(n)\propto\sqrt{nB}$ at low energies. Also notice
that this method could not account for the complicated relation
between $E^c$ and ($n,B$) at high energies, and the important
differences in state degeneracy and level spacing at different
energy regimes.

Density of states, which directly reflects the main
characteristics of Landau levels, is
\begin{align}
D(\omega)=\sum_{\sigma,h=c,\nu}\int_{1st BZ}
\frac{dk_y}{2\pi}\frac{\Gamma}{\pi}\frac{1}{[\omega-E^{h}(k_y)]^2+\Gamma^2},
\end{align}
where the broadening factor is $\Gamma=10^{-3}\gamma_0$.
$D(\omega)$ without B, as shown by the dashed curve in Figs.
4(a)-4(c), exhibits a vanishing value at $E_F=0$, the linear
$\omega$-dependence at low frequency, and one symmetric
logarithmic divergence at $\omega=\gamma_0$. These features are
drastically changed by the magnetic field. $D(\omega)$ is finite
at $E_F=0$ and exhibits a lot of delta-function-like peaks at
$\omega\neq0$. Such prominent peaks come from the 0D Landau
levels. The peak height represents the Landau-level degeneracy, so
peaks at $\omega<\gamma_0$ [Fig. 4(a)] are higher than those at
$\omega>\gamma_0$ [Fig. 4(c)]. The distribution of peaks is
nonuniform because of the unequally spaced Landau levels. The
Landau levels with $E^c(n)\sim\gamma_0$ are oscillatory and dense
so that the prominent peaks might be replaced by the weak
broadening peaks [Fig. 4(b)]. Density of states is closely related
to the available channels of optical excitations. The predicted
features could be examined by the optical spectroscopy.

In conclusion, the magneto-electronic properties are calculated by
the Peierl's tight-binding model with the band Hamiltonian matrix.
The magnetic field could make linear and parabolic bands become
Landau levels. Most of Landau levels are dispersionless, with some
oscillatory Landau levels at $E^c\sim\gamma_0$ excepted. State
degeneracy is, respectively, double and fourfold for
$E^c(n)>\gamma_0$ and $E^c(n)<\gamma_0$. The Landau-level spacing
decreases sharply at low energies, while it increases slowly at
high energies. The Landau-level energies are characterized by a
simple relation $E^c(n)\propto\sqrt{nB}$ only for
$E^c(n)\lesssim0.4\gamma_0$, not for others. Density of states
exhibits many delta-function-like prominent peaks, mainly owing to
the 0D Landau levels. The effective-mass method can not account
for energy dispersion, state degeneracy, level spacing, and
dependence of state energy on ($n,B$). The magneto-optical
absorption spectra could be used to verify the predicted
electronic properties.

This work was supported by the National Science Council of Taiwan,
under the Grant Nos. NSC 95-2112-M-006-002.

\newpage

\begin{figure}[p]
%\hspace*{-5in}
\begin{center}
\includegraphics[scale=0.7]{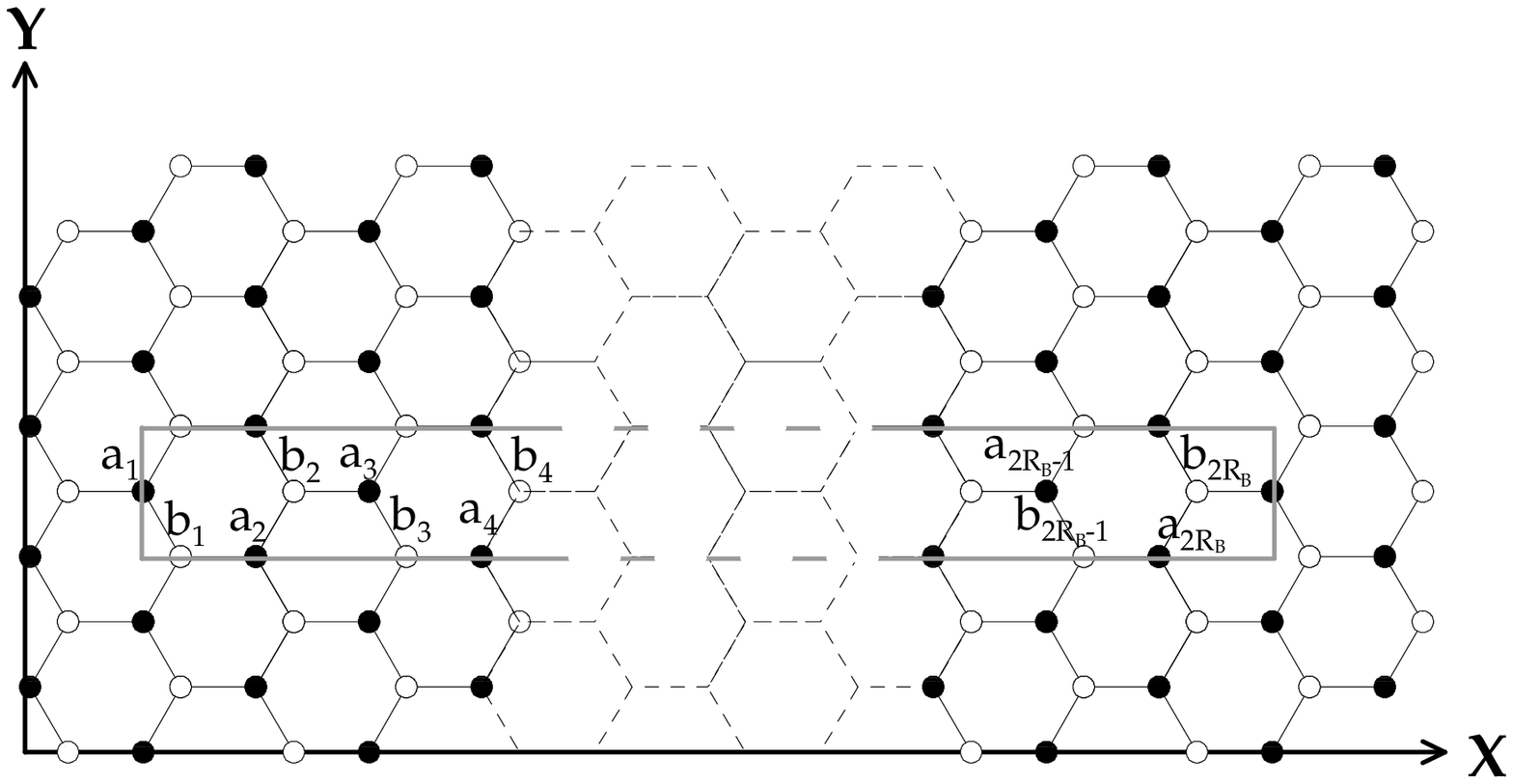}
%\vspace*{-1.5in}
\caption{The primitive unit cell of a monolayer graphite in a
perpendicular uniform magnetic field.}
\end{center}
\end{figure}

\begin{figure}[p]
%\hspace*{-5in}
\begin{center}
\includegraphics[scale=0.7]{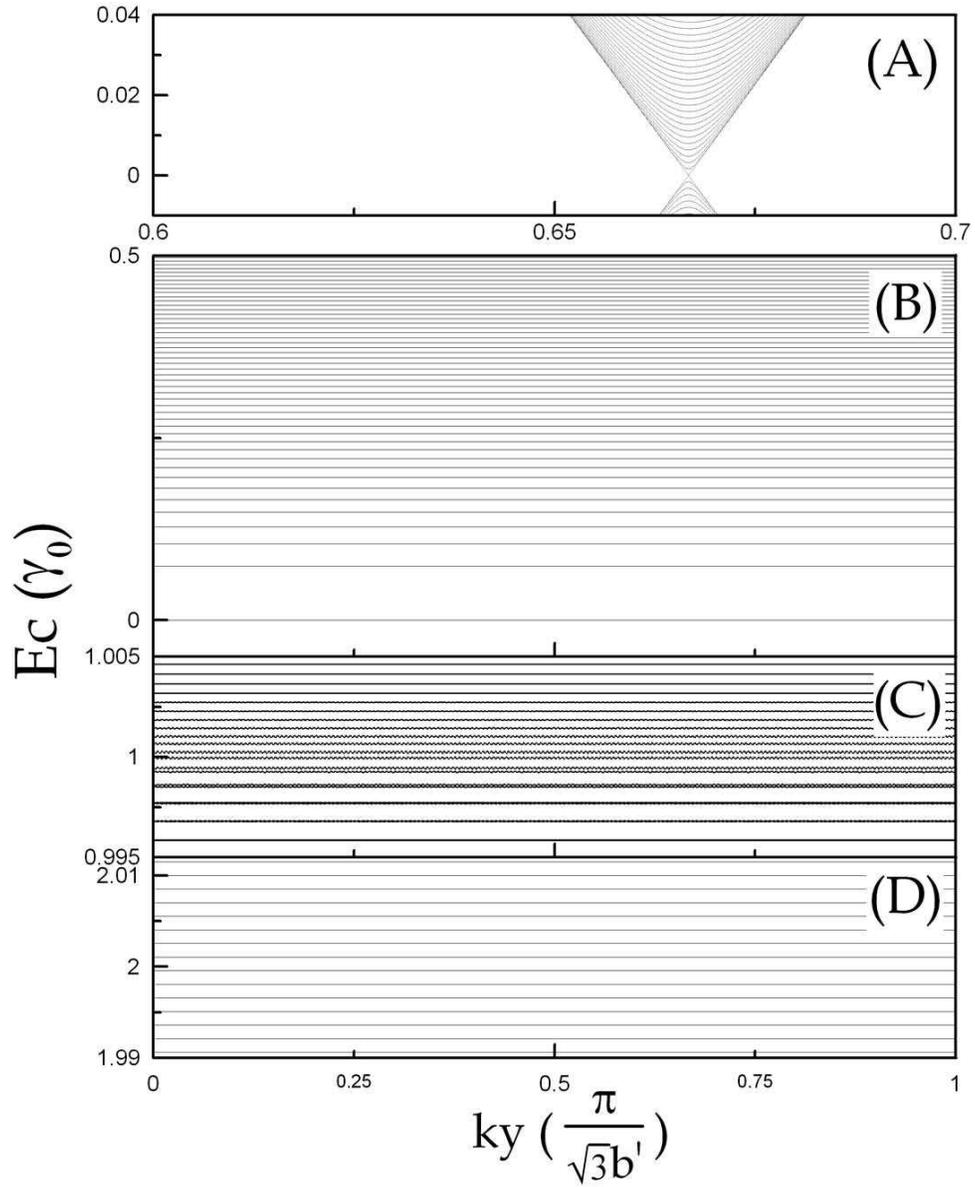}
%\vspace*{-1.5in}
\caption{Landau levels of B=40 T at different energy regimes (b),
(c), and (d). The low energy bands of B=0 are shown in (a).}
\end{center}
\end{figure}

\begin{figure}[p]
\begin{center}
\includegraphics[scale=0.7]{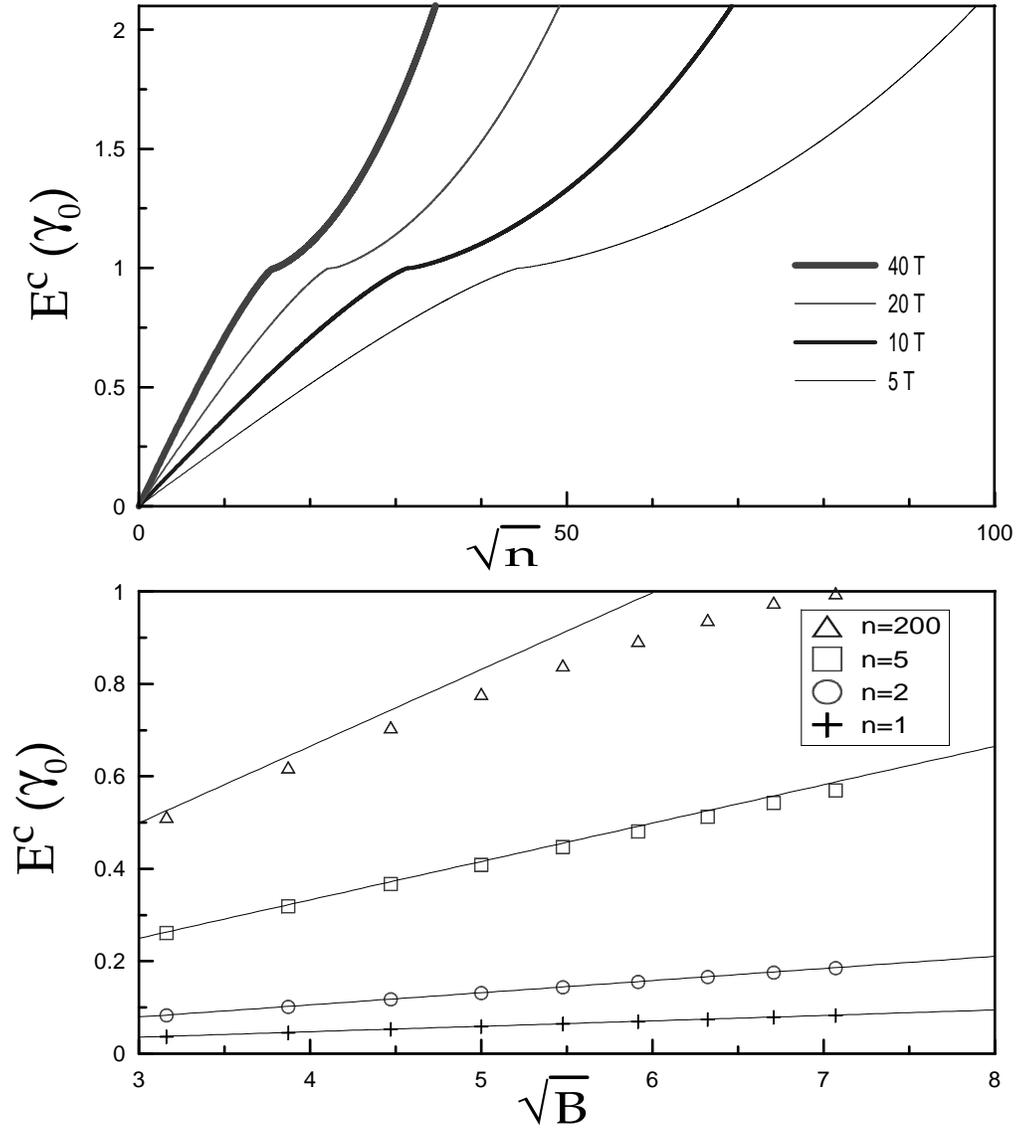}
\caption{The variation of low Landau-level energies with square
root of (a)quantum number and (b)magnetic field. The solid curves
in (b) are the linear guidelines.}
\end{center}
\end{figure}

\begin{figure}[p]
\begin{center}
\includegraphics[scale=0.7]{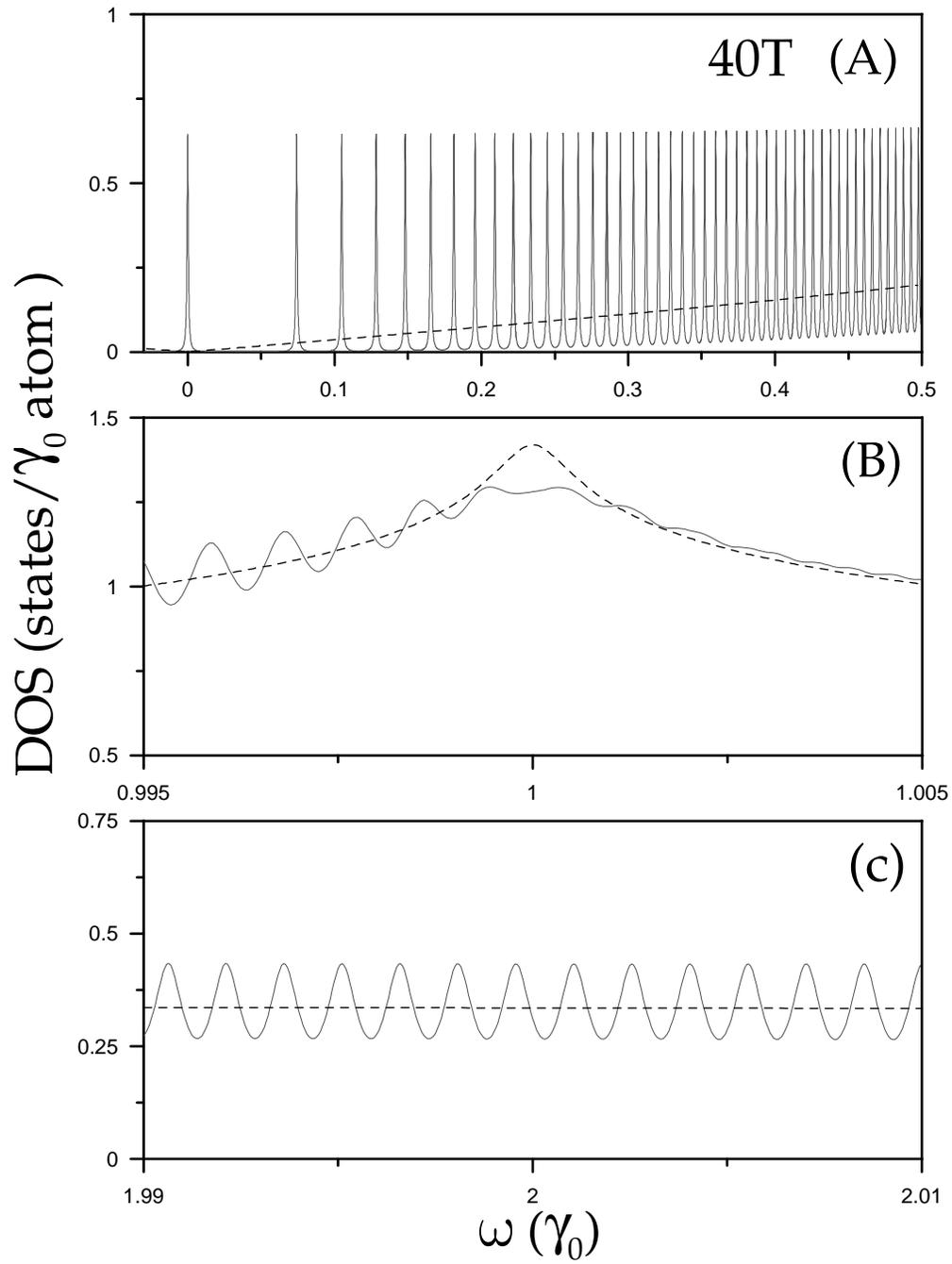}
\caption{Density of states of B=40 T in different energy regimes
(a), (b), and (c). That at B=0 is also shown by the dashed curve.}
\end{center}
\end{figure}
\end{document}